\documentclass[]{aa}  

\usepackage{graphicx}
\usepackage{txfonts}
\usepackage{orcidlink}
\usepackage{hyperref}     
\hypersetup{colorlinks=true,linkcolor=blue,citecolor=blue,filecolor=blue,urlcolor=blue}

\def \kms {{\rm km\,s$^{-1}$}}

\def \msun {{$\mathrm{M}_\odot$}}

\begin{document} 

\title{Cracking open the `dragon egg'}
\subtitle{ALMA 1.3\,mm observations of a compact, young high-mass core candidate at 30\,mas resolution}

\authorrunning{A.~T.~Barnes et al. }

\author{
    A.~T.~Barnes \inst{\ref{eso}}\orcidlink{0000-0003-0410-4504}\thanks{\url{ashleybarnes.astro@gmail.com}} \and 
    J.~D.~Henshaw \inst{\ref{mpia}}\orcidlink{0000-0001-9656-7682} \and
    \'A.~S\'anchez-Monge \inst{\ref{ice},\ref{ieec}}\orcidlink{0000-0002-3078-9482} \and
    P.~Caselli \inst{\ref{mpe}} \and
    I.~Jim\'enez-Serra \inst{\ref{cab_csic}} \and
    J.~E.~Pineda \inst{\ref{mpe}}\orcidlink{0000-0002-3972-1978}  \and
    K.~Immer \inst{\ref{eso}} \and
    M.~De~Simone \inst{\ref{eso}} \and
    R.~J.~Parker \inst{\ref{ushef}}\thanks{Royal Society Dorothy Hodgkin Fellow} \and
    F.~Fontani \inst{\ref{inaf}} \and
    E.~Redaelli \inst{\ref{eso}}\orcidlink{0000-0002-0528-8125} \and
    S.~N.~Longmore \inst{\ref{ljmu}} \and
    L.~B\"ohm \inst{\ref{eso}} \and
    Q.~Zhang \inst{\ref{cfa}} \and 
    C.~Y.~Law \inst{\ref{inaf}} \and
    G.~Cosentino \inst{\ref{iram}} \and 
    J.~C.~Tan \inst{\ref{chalmers}} 
    }

\institute{
    \label{eso}{European Southern Observatory (ESO), Karl-Schwarzschild-Stra{\ss}e 2, 85748 Garching, Germany}  \and
    \label{mpia}{Max-Planck-Institut für Astronomie, Heidelberg, Germany} \and
    \label{ice}{Institut de Ci\`encies de l'Espai (ICE), CSIC, Campus UAB, Carrer de Can Magrans s/n, E-08193, Bellaterra, Barcelona, Spain} \and
    \label{ieec}{Institut d'Estudis Espacials de Catalunya (IEEC), E-08860, Castelldefels, Barcelona, Spain} \and
    \label{mpe}{Max-Planck-Institut f\"{u}r extraterrestrische Physik, Giessenbachstrasse 1, 85748 Garching bei M\"{u}nchen, Germany} \and
    \label{cab_csic}{Centro de Astrobiolog\'ia (CAB), CSIC-INTA, Ctra. de Ajalvir Km. 4, 28850, Torrej\'on de Ardoz, Madrid, Spain} \and
    \label{ushef}{Astrophysics Research Cluster, School of Mathematical and Physical Sciences, The University of Sheffield, Hounsfield Road, Sheffield S3 7RH, UK} \and
    \label{inaf}{INAF Osservatorio Astrofisico di Arcetri, Largo E. Fermi 5, 50125 Florence, Italy} \and
    \label{ljmu}{Astrophysics Research Institute, Liverpool John Moores University, 146 Brownlow Hill, Liverpool L3 5RF, UK} \and
    \label{cfa}{Center for Astrophysics | Harvard \& Smithsonian, Cambridge, MA, USA} \and
    \label{chalmers}{Dept. of Space, Earth and Environment, Chalmers University of Technology, SE-412 96 Gothenburg, Sweden} \and 
    \label{iram}{Institut de Radioastronomie Millimétrique, 300 Rue de la Piscine, 38400 Saint-Martin-d’Hères, France}
}

\date{Received 20/04/2025; accepted 22/09/2026}

\abstract{
Massive pre-stellar cores are a central assumption of core-accretion models of high-mass star formation, yet convincing observational examples remain rare, with recent surveys generally finding widespread fragmentation in massive clumps. The `dragon egg' (C2c1a) core in the infrared-dark cloud G028.37+00.07 (the `dragon cloud') has been proposed as {a promising compact high-mass core candidate}. We present new ALMA Band~6 observations at $0.035''$--$0.078''$ ($\sim175$--400\,au at 5\,kpc) of C2c1, providing one of the most stringent high-resolution tests to date of a compact massive core candidate. The region resolves into multiple fragments, but the brightest component, C2c1a(a), remains a single compact continuum peak down to $\sim175$\,au, with a deconvolved radius of $\sim87$\,au {and a nearest neighbouring fragment at a projected separation of $\sim2400$\,au}. The continuum peak reaches a brightness temperature of $\sim40$\,K.\ This significantly exceeds the $\sim10$\,K gas temperatures inferred from lower-resolution NH$_3$ observations, indicating a strong radial temperature gradient. Continuum-based mass estimates for the compact component range from {$\sim2$\,\msun} to $\sim30$\,\msun\ (implying extreme densities of $n_\mathrm{H2} > 10^9$\,cm$^{-3}$) depending on the assumed temperature, while lower-resolution data indicate that it is embedded within a substantially larger mass reservoir of $\sim110$\,\msun\ on $\sim10^4$\,au scales. The spectrum towards the continuum peak shows no strong hot-core tracers and only weak, spatially extended H$_2$CO and CO emission; no high-velocity outflow can be unambiguously associated with C2c1a(a). These properties are consistent with a very early evolutionary stage. We conclude that C2c1a(a) is {an exceptionally compact core embedded within a massive cold reservoir, with no additional compact fragments detected within the present resolution and sensitivity limits}. The data are also consistent with a strongly stratified structure in which a compact, warm inner component is embedded within a colder and more massive envelope, potentially tracing the earliest formation of {a central protostellar `seed' within a still-dominant reservoir}.
}
\maketitle
\nolinenumbers

\section{Introduction}
\label{sec:intro}

\begin{figure*}[!t]
\centering
\includegraphics[width=\textwidth]{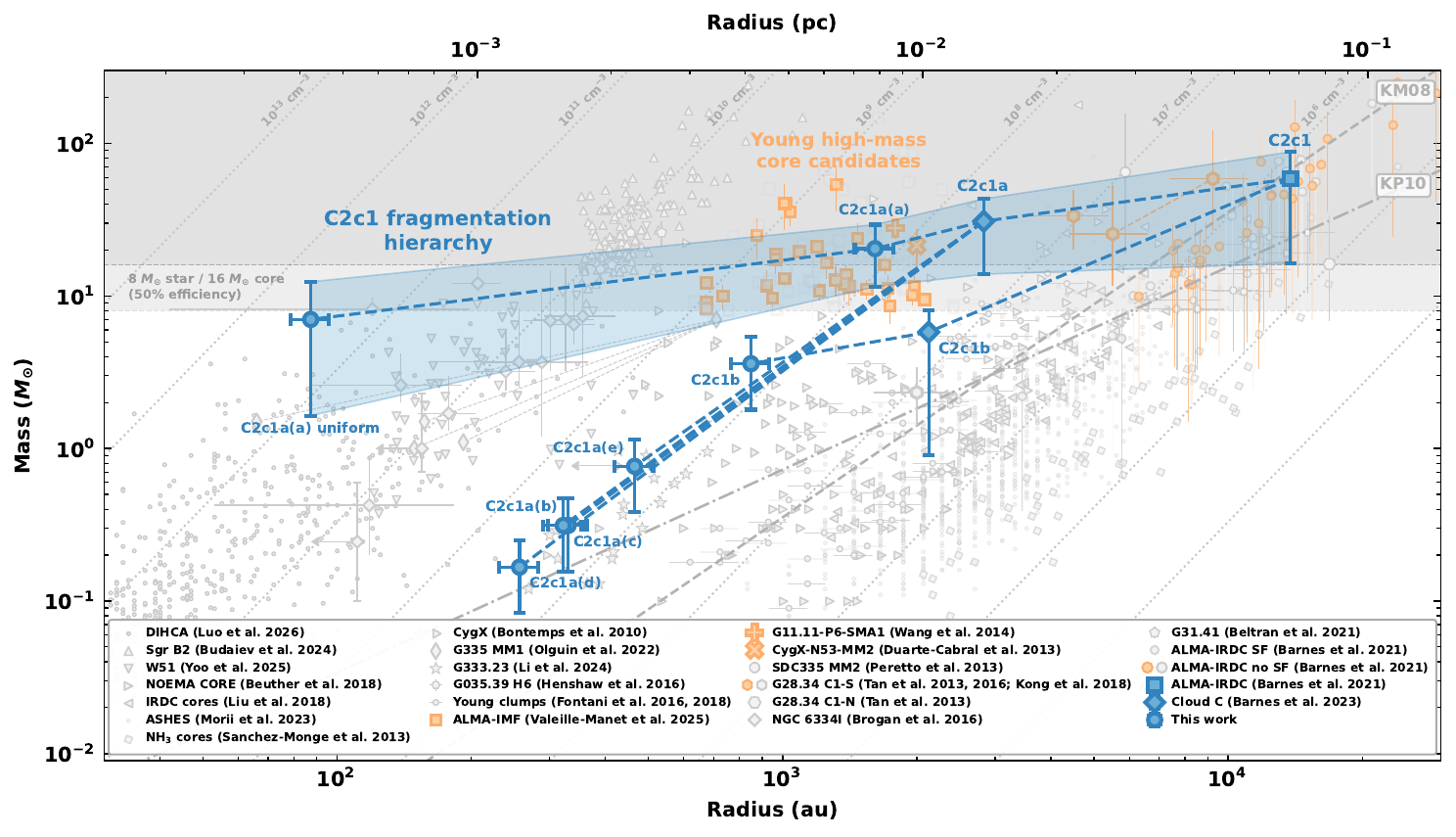} \vspace{-5mm}
\caption{
{Core mass as a function of radius for compact cores and condensations in high-mass star-forming environments. Effective radii are shown except for C2c1a(a) in the uniformly weighted image, for which we use the deconvolved radius. Grey symbols show the broader literature compilation, which draws on \citet{Bontemps2010}, \citet{SanchezMonge2013}, \citet{Henshaw2016a}, \citet{Brogan2016}, \citet{Fontani2016,Fontani2018}, \citet{Beuther2018}, \citet{Liu2018}, \citet{Tan2013,Tan2016}, \citet{Kong2018}, \citet{Wang2014}, \citet{Duarte-Cabral2013}, \citet{Peretto2013}, \citet{Beltran2021}, \citet{Olguin2022}, \citet{Morii2023}, \citet{Budaiev2024}, \citet{Li2024b}, \citet{Valeille-Manet2025}, \citet{Yoo2025}, \citet{Luo2026}, and \citet{Barnes2021c}. Orange symbols highlight comparatively young high-mass core candidates: the ALMA-IMF pre-stellar-core candidates \citep{Valeille-Manet2025}, G11.11$-$P6$-$SMA1 \citep{Wang2014}, CygX$-$N53$-$MM2 \citep{Duarte-Cabral2013}, G28.34 C1-S \citep{Tan2013,Tan2016,Kong2018}, and ALMA-IRDC cores without identified star-formation indicators \citep{Barnes2021c}. Blue symbols and dashed connecting lines trace the hierarchy of Cloud~C from C2c1 \citep{Barnes2021c}, through C2c1a and C2c1b \citep{Barnes2023b}, to the six components identified in this work (see Sect.~\ref{sec:continuum}). The horizontal dashed lines mark an $8\,M_\odot$ star and the $16\,M_\odot$ core required to form such a star for a core-to-star efficiency of 50\%. The dashed and dot-dashed grey curves show the high-mass star-formation thresholds of \citet{Krumholz2008} and \citet{Kauffmann2010}, respectively, while the dotted diagonals indicate constant molecular-hydrogen number density. Appendix~\ref{sec:app_literature_comparison} describes the sample selection, radius definitions, mass assumptions, and derived quantities, which we also provide in a machine-readable format.}
}
\label{fig:lit_comparison}
\end{figure*}

Massive stars ($M \gtrsim 8\,M_\odot$) strongly influence the structure and evolution of galaxies through ionising radiation, stellar winds, and supernova feedback \citep[e.g.][]{Krumholz2014}. How they form, however, remains debated. {Their formation begins in the densest parts of molecular clouds, but the physical state of these regions before a luminous (and disruptive) high-mass protostar becomes visible remains poorly constrained. This earliest phase is expected to be short-lived, deeply embedded, and strongly affected by fragmentation, accretion from larger scales, and radiative feedback once a central source forms (see reviews \citealp{Tan2014, Motte2018a, Beuther2025a}).} {Several theoretical frameworks describe these processes in different ways, from models in which a massive bound core is assembled early \citep[e.g.][]{McLaughlin1997, Mckee2003} to scenarios in which compact protostellar seeds continue to grow through accretion from a larger-scale clump potential \citep[e.g.][]{Bonnell2001a, Wang2010, Peretto2013}. Additional variants emphasise global hierarchical collapse, hub--filament accretion flows, and time-dependent fragmentation within dynamically evolving clumps (see e.g. \citealp{Vazquez-Semadeni2019, Padoan2020}).} {A direct observational route through this problem is to identify the youngest, most compact, and most massive condensations possible, determine whether they already host embedded star formation activity, and assess their physical properties to constrain the models.}

{Figure\,\ref{fig:lit_comparison} places the search for young high-mass cores within a broad literature compilation of high-angular-resolution observations of high-mass star-forming regions.\footnote{{The compilation is not intended to provide a complete or unbiased census. The contributing studies differ in their target selection, completeness, sensitivity, angular resolution, spatial filtering, observing facility, and methods used to identify and characterise compact sources. The resulting distribution should therefore be regarded as representative of the available measurements rather than as a statistically homogeneous sample. See Appendix~\ref{sec:app_literature_comparison} for a description of the sample selection, radius definitions, mass assumptions, and derived quantities.}} It spans sources classified as starless, pre-stellar, and protostellar and includes compact condensations embedded in more evolved high-mass star-forming environments. Within this diverse population, only a small subset appears sufficiently massive to form a high-mass star (typically $M\gtrsim16\,M_\odot$ for a core-to-star efficiency of approximately 50\%), compact, and free from clear evidence of embedded activity (highlighted in Fig~\ref{fig:lit_comparison}). Prominent candidates include G28.37$-$C1-S \citep{Tan2013,Tan2016,Kong2018}, CygX$-$N53$-$MM2 \citep{Duarte-Cabral2013}, G11.11$-$P6$-$SMA1 \citep{Wang2014}, and W43$-$MM1~\#6 \citep{Nony2018,Valeille-Manet2025}. However, the interpretation of these candidates can still be complicated by nearby protostellar activity, indications of internal heating, or evidence of embedded star formation activity revealed by subsequent observations \citep[e.g.][]{Cyganowski2022}. The small number of sources satisfying these criteria may therefore reflect a combination of observational incompleteness for cold, compact cores and the intrinsically short interval between the assembly of a massive core and the emergence of detectable signs of embedded star formation.}

\begin{figure*}
\centering
\includegraphics[width=\textwidth]{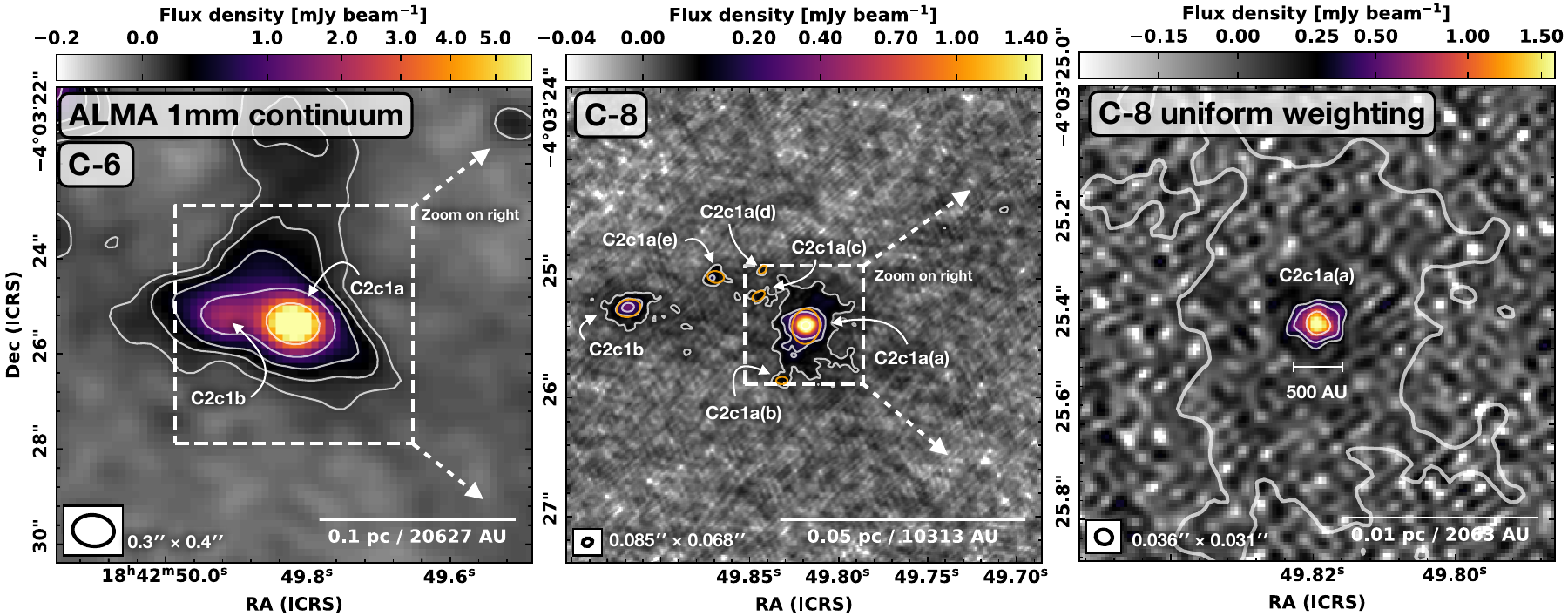}
\caption{ALMA 1\,mm continuum maps of the C2c1 region at increasing angular resolution. \textit{Left}: C-6 configuration image, tracing the large-scale continuum morphology and revealing two dominant fragments embedded within diffuse emission \citep[see][]{Barnes2023b}. {Contours are shown at 0.07, 0.19, 0.37, 1.12, and 3.72\,mJy\,beam$^{-1}$, corresponding to 2, 5, 10, 30, and 100\,$\sigma$.}
\textit{Middle}: C-8 naturally weighted image, resolving the two main cores and the lower-level connecting emission. Overlaid are the positions of cores identified from the dendrogram analysis. {Contours are shown at 0.05, 0.13, and 0.25\,mJy\,beam$^{-1}$, corresponding to 2, 5, and 10\,$\sigma$.}
\textit{Right}: C-8 uniformly weighted image, isolating the compact emission associated with the primary core. {Contours are shown at 0.28, 0.47, and 0.95\,mJy\,beam$^{-1}$, corresponding to 3, 5, and 10\,$\sigma$. For reference, the thick white contour indicates the lowest contour level from the naturally weighted image shown in the middle panel.} In the uniformly weighted image, the primary core appears unresolved or only marginally resolved and is characterised by a single compact peak at the present sensitivity, with no statistically significant sub-fragmentation down to $\sim175$\,au.
Squares overlaid on the left and middle panels indicate the region displayed in the subsequent higher-resolution panel, illustrating the progressive zoom from left to right. Synthesised beams are shown in the lower-left corner of each panel, and physical scale bars are indicated.
}
\label{fig:continuum_maps}
\end{figure*}

{One particularly promising target for such a high-resolution test is C2c1a,} the central fragment of the C2c1 region in G028.37+00.07 (the `dragon cloud' or Cloud~C; \citealp{Butler2009, Wang2015}). The C2 region is associated with a massive Herschel infrared Galactic Plane Survey (Hi-GAL) clump ($M=1061\,M_\odot$) that has a low luminosity-to-mass ratio, $L/M\simeq0.5\,L_\odot\,M_\odot^{-1}$ \citep{Elia2021}, indicative of a very early stage of high-mass star formation. Within this clump, Atacama Large Millimeter/submillimeter Array (ALMA) observations identified the C2c1 region, a compact ($\sim110\,M_\odot$ within an effective radius of $\sim10\,000$\,au; see Fig.~\ref{fig:lit_comparison}) core candidate that remains exceptionally cold and quiescent \citep{Barnes2021c}. In particular, the central fragment, C2c1a (the `dragon egg'), was inferred to contain $20$--$30\,M_\odot$ within an effective radius of only $\sim2800$\,au, while exhibiting strong deuteration and CO depletion, but no clear evidence of an associated molecular outflow \citep{Zhang2015, Kong2019, Barnes2016, Barnes2023b}. On larger scales, G028.37+00.07 has recently been characterised as part of a dynamically evolving converging gas flow with substantial longitudinal mass transport and complex infall motions towards the core region \citep{Beuther2026}. Together, these results identify C2c1a as an exceptional laboratory for studying the earliest stages of high-mass star formation. However, the currently available observations lack the angular resolution needed to determine whether the emission remains dominated by a single compact component, whether the inferred mass is concentrated on sub-1000\,au scales, and whether the source has already formed a central protostar.

This work presents ALMA Band~6 observations at $0.035''$--$0.08''$ resolution ($\sim\,175$--400\,au at 5\,kpc; \citealp{Simon2006}), providing the highest-resolution view of the dragon egg region to date (see Fig.\,\ref{fig:continuum_maps}). {These data are among the highest-resolution and most sensitive ALMA Band~6 observations currently available for similar targets (see Fig.\,\ref{fig:lit_comparison_alma}). We use them to examine C2c1a (dragon egg) as a possible snapshot of the initial stages of high-mass star formation and to determine: (i) whether the compact peak contains enough mass to remain relevant for high-mass star formation, (ii) whether its emission is concentrated into one very compact component or breaks into multiple fragments, (iii) whether its chemistry remains consistent with a very young source, and (iv) whether any outflow can be directly associated with the compact peak.} 

Section~\ref{sec:obs} describes the observation strategy. Section~\ref{sec:continuum} presents the continuum morphology and compactness constraints. Section~\ref{sec:mass_constraints} discusses the mass estimates and associated uncertainties. Section~\ref{sec:chemistry} evaluates the kinematic and chemical diagnostics, and assesses the outflow evidence. Section~\ref{sec:discussion} {interprets C2c1a(a) as an exceptionally young and compact high-mass core candidate} in light of these results, and Sect.~\ref{sec:conclusions} summarises the main conclusions.

\section{Observations}
\label{sec:obs}

We observed the C2c1 region with ALMA in Band~6 (1.3\,mm) as part of project \mbox{2025.1.00675.S}. In this work, we use only the most extended (C-8; 110 to 8500\,m baseline range) configuration data from this programme (at the time of publication, only these data were available), which provide high angular resolution on spatial scales of a few hundred au.
{Given the technical challenge of such high-angular-resolution observations, the correlator setup was designed primarily to maximise continuum sensitivity, while retaining sufficient spectral sensitivity and resolution to detect and characterise only the brightest molecular lines.}

The correlator setup for these observations is summarised in Appendix\,\ref{sec:app_spectral_setup} (see Table~\ref{tab:spectral_setup}). Four narrowband spectral windows target the molecular transitions N$_2$D$^+$\,(3--2), CO\,(2--1), $^{13}$CS\,(5--4), and OCS\,(19--18), each with a bandwidth of 58.6\,MHz and native velocity resolutions of 0.16--0.18\,km\,s$^{-1}$. These tracers probe cold dense gas and outflow activity, enabling an assessment of the internal kinematics of the core. In addition, three wideband spectral windows were configured for continuum observations, each with a bandwidth of 1.875\,GHz and an effective velocity resolution of $\sim1.5$\,km\,s$^{-1}$. These windows also encompass several molecular transitions, including SiO\,(5--4) and weaker lines (see \citealp{Barnes2023b}). Combining the three continuum windows and excluding channels affected by line emission yields an effective continuum bandwidth of 2.95\,GHz.

Calibration and imaging were performed using the ALMA pipeline (\textsc{casa} version 6.6.6-17, pipeline version 2025.1.0.35; \citealp{CASATeam2022, Hunter2023}). We imaged the data using Briggs weighting with a robust parameter of $+1$, which we refer to as `fiducial' weighting. The angular resolution varies slightly between the individual spectral windows, ranging from $0.0891'' \times 0.0704''$ to $0.0787'' \times 0.0659''$. This corresponds to a characteristic angular resolution of $\sim0.08''$ ($\sim400$\,au, or $\sim0.002$\,pc, at 5\,kpc). For the narrowband spectral windows, the typical rms noise level in the fiducially weighted images is $\sim1.5$\,mJy\,beam$^{-1}$ ($\sim6$\,K) per channel. For the individual wideband spectral windows, the rms noise level is $\sim0.5$\,mJy\,beam$^{-1}$ ($\sim2$\,K) per channel. The continuum image has an rms noise level of 0.017\,mJy\,beam$^{-1}$.

To probe smaller spatial scales, we additionally imaged the continuum data using Briggs weighting with a robust parameter of $-2$ (approximately uniform weighting), trading sensitivity for angular resolution. In this case, we combined the full 5.625\,GHz bandwidth of the three continuum windows, assuming negligible line contamination (see Fig.~\ref{fig:aperture_spectra}). The resulting continuum image has a restoring beam of $0.0358'' \times 0.0313''$ at a position angle of $71.9\degr$ and an rms noise level of 0.095\,mJy\,beam$^{-1}$. This corresponds to a characteristic angular resolution of $\sim0.035''$ ($\sim175$\,au).

\begin{table*}
\centering
\caption{Dendrogram-derived properties measured from the fiducial (robust parameter of +1) weighted image ($\sim0.08\arcsec$ resolution; see Sect.~\ref{sec:obs}).}
\label{tab:dendrogram_props}
\begin{tabular}{lcccccc}
\hline\hline
Property & C2c1a(a) & C2c1b & C2c1a(e) & C2c1a(b) & C2c1a(c) & C2c1a(d) \\
\hline
$\alpha_{\mathrm{J2000}}$ (deg) & 280.707577 & 280.707992 & 280.707788 & 280.707688 & 280.707633 & 280.707678 \\
$\delta_{\mathrm{J2000}}$ (deg) & -4.057055 & -4.057013 & -4.056940 & -4.056985 & -4.057182 & -4.056923 \\
$S_{\nu}$ (mJy) & $11.0\pm1.1$ & $1.50\pm0.15$ & $0.33\pm0.03$ & $0.13\pm0.01$ & $0.13\pm0.01$ & $0.071\pm0.007$ \\
$S_{\nu,\mathrm{peak}}$ (mJy) & $3.8\pm0.4$ & $0.37\pm0.04$ & $0.16\pm0.02$ & $0.090\pm0.009$ & $0.087\pm0.009$ & $0.068\pm0.007$ \\
$T_{\mathrm{b,max}}$ (K) & $16.5\pm1.6$ & $1.6\pm0.2$ & $0.7\pm0.1$ & $0.4\pm0.1$ & $0.4\pm0.1$ & $0.3\pm0.1$ \\
$r$ (arcsec) & $0.111\pm0.011$ & $0.073\pm0.007$ & $0.050\pm0.005$ & $0.039\pm0.004$ & $0.033\pm0.003$ & $0.027\pm0.003$ \\
$r_{\mathrm{eff}}$ (arcsec) & $0.322\pm0.032$ & $0.170\pm0.017$ & $0.093\pm0.009$ & $0.066\pm0.007$ & $0.064\pm0.006$ & $0.051\pm0.005$ \\
$s$ (arcsec) & -- & $1.50\pm0.15$ & $0.86\pm0.09$ & $0.47\pm0.05$ & $0.50\pm0.05$ & $0.60\pm0.06$ \\
$r$ (AU) & $560\pm60$ & $370\pm40$ & $250\pm30$ & $200\pm20$ & $160\pm20$ & $140\pm14$ \\
$r_{\mathrm{eff}}$ (AU) & $1600\pm160$ & $850\pm90$ & $470\pm50$ & $330\pm30$ & $320\pm30$ & $260\pm30$ \\
$r_{\mathrm{deconv}}$ (AU) & $520\pm50$ & $320\pm30$ & $170\pm17$ & $77\pm8$ & -- & -- \\
$r_{\mathrm{eff,deconv}}$ (AU) & $1600\pm160$ & $830\pm80$ & $430\pm40$ & $280\pm30$ & $270\pm30$ & $180\pm20$ \\
$s$ (AU) & -- & $7500\pm700$ & $4300\pm400$ & $2400\pm200$ & $2500\pm200$ & $3000\pm300$ \\
$M_{10\mathrm{K}}$ ($M_{\odot}$) & $26.7\pm2.7$ & $3.6\pm0.4$ & $0.76\pm0.08$ & $0.31\pm0.03$ & $0.31\pm0.03$ & $0.17\pm0.02$ \\
$M_{T_{\mathrm{b,max}}}$ ($M_{\odot}$) & $12.8\pm1.3$ & -- & -- & -- & -- & -- \\
$n_{10\mathrm{K}}$ ($10^8$ cm$^{-3}$) & $48\pm15$ & $22\pm7$ & $15\pm5$ & $13\pm4$ & $22\pm7$ & $20\pm6$ \\
$n_{T_{\mathrm{b,max}}}$ ($10^8$ cm$^{-3}$) & $23\pm7$ & -- & -- & -- & -- & --
\smallskip \\
\hline
\end{tabular}
\tablefoot{
    The coordinates ($\alpha_{\mathrm{J2000}}$, $\delta_{\mathrm{J2000}}$) give the intensity-weighted centroid positions of each fragment. $S_{\nu}$ and $S_{\nu,\mathrm{peak}}$ denote the integrated and peak continuum flux densities, respectively, and $T_{\mathrm{b,max}}$ is the corresponding peak brightness temperature. The radius $r$ (arcsec) is derived from the two-dimensional second moment of the emission, while the effective radius $r_{\mathrm{eff}}$ is computed from the projected area $A$ of the dendrogram-identified structure assuming circular symmetry, $r_{\mathrm{eff}}=\sqrt{A/\pi}$. Physical radii are obtained assuming a distance of $d=5$\,kpc \citep{Simon2006}. Deconvolved radii are obtained by subtracting the synthesised beam in quadrature and are reported only when the source is spatially resolved {(i.e. moment size estimates larger than the beam size)}; otherwise entries are marked `--'. The quantity $s$ denotes the projected separation from C2c1a(a). Uncertainties correspond to $1\sigma$ values and include an assumed $\sim10$\% systematic uncertainty in the absolute ALMA flux calibration. A fractional uncertainty of 10\% is adopted for size-related quantities. Masses and densities are derived as described in Sect.~\ref{sec:mass_constraints}, assuming optically thin emission and adopting either $T_{\mathrm{dust}}=10$\,K or $T_{\mathrm{dust}}=T_{\mathrm{b,max}}$ {(although we may be missing a significant fraction of the flux if the dust emission is optically thick; see Sect.~\ref{sec:mass_constraints}).} Densities assume a spherically symmetric core geometry. For fragments other than C2c1a(a), the measured $T_{\mathrm{b,max}}$ values are very low and would formally imply unrealistically high masses and densities; this indicates that the emission is optically thin and that $T_{\mathrm{b,max}}$ does not represent the true dust temperature. We therefore report $T_{\mathrm{b,max}}$-based masses and densities only for C2c1a(a). Combining uncertainties in flux calibration, dust opacity ($\sim30$\%; \citealp{Sanhueza2017}), and distance results in an overall systematic uncertainty of approximately 50\%.
}
\end{table*}

\begin{table}
\centering
\caption{Dendrogram-derived properties of C2c1a(a) measured from the uniform weighted image ($\sim0.035\arcsec$ resolution; see Sect.~\ref{sec:obs}).}
\label{tab:dendrogram_props_uni}
\begin{tabular}{lc}
\hline\hline
Property & C2c1a(a) (uniform weighted) \\
\hline
$\alpha_{\mathrm{J2000}}$ (deg) & 280.707581 \\
$\delta_{\mathrm{J2000}}$ (deg) & -4.057052 \\
$S_{\nu}$ (mJy) & $4.9\pm0.6$ \\
$S_{\nu,\mathrm{peak}}$ (mJy) & $1.8\pm0.2$ \\
$T_{\mathrm{b,max}}$ (K) & $39\pm4$ \\
$r$ (arcsec) & $0.023\pm0.002$ \\
$r_{\mathrm{eff}}$ (arcsec) & $0.054\pm0.005$ \\
$r$ (AU) & $120\pm12$ \\
$r_{\mathrm{eff}}$ (AU) & $270\pm30$ \\
$r_{\mathrm{deconv}}$ (AU) & $87\pm9$ \\
$r_{\mathrm{eff,deconv}}$ (AU) & $260\pm30$ \\
$M_{10\mathrm{K}}$ ($M_{\odot}$) & $11.2\pm1.3$ \\
$M_{T_{\mathrm{b,max}}}$ ($M_{\odot}$) & $1.9\pm0.2$ \\
$n_{10\mathrm{K}}$ ($10^8$ cm$^{-3}$) & $2200\pm700$ \\
$n_{T_{\mathrm{b,max}}}$ ($10^8$ cm$^{-3}$) & $360\pm120$ \smallskip \\
\hline
\end{tabular}
\tablefoot{
    See the Table\,\ref{tab:dendrogram_props} footnotes.}
\end{table}

\section{Fragmentation structure}
\label{sec:continuum}

\begin{figure}
\centering
\includegraphics[width=\columnwidth]{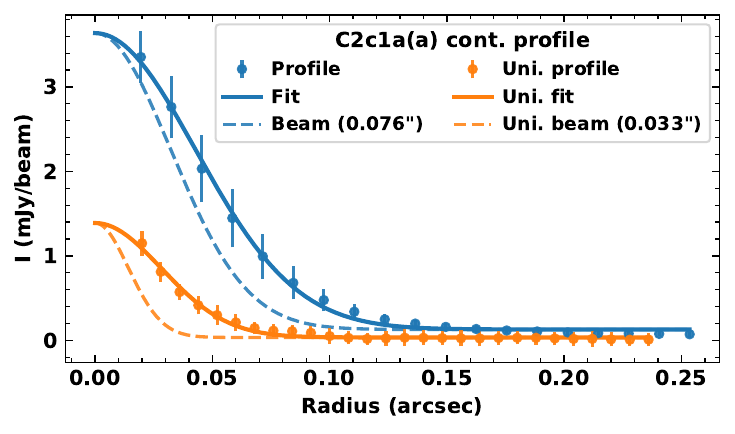}
\caption{Azimuthally averaged 1\,mm continuum radial intensity profiles of C2c1a(a) for the naturally weighted (blue) and uniformly weighted (orange) images. 
    Points with error bars show the mean intensity and scatter in concentric annuli, and solid curves show the best-fit Gaussian-plus-constant models with the Gaussian centroid fixed at $r=0$. Dashed curves show the corresponding (circularised; i.e. $\sqrt{\theta_\mathrm{maj}\theta_\mathrm{min}}$, where $\theta_\mathrm{maj}$ and $\theta_\mathrm{min}$ are the major and minor Gaussian beam axes) synthesised-beam Gaussian profiles, scaled to the fitted peak intensity; the beam FWHM values are listed in the legend. 
    The uniform-weighted profile is narrower and has lower total recovered flux, consistent with enhanced filtering of extended emission at higher angular resolution. Intensity is shown in mJy\,beam$^{-1}$ and radius in arcsec.
    }
\label{fig:radial_profile}
\end{figure}

At the spatial scales probed by the new C-8 data, the C2c1 region resolves into two compact continuum cores, C2c1a and C2c1b {separated by $\sim7500$\,au}, which are embedded within the more extended emission seen in the lower-resolution C-6 image (Fig.~\ref{fig:continuum_maps}; \citealp{Barnes2023b}). Previous observations, at the same frequency (1.3\,mm), identified C2c1a as a single core with an effective radius of $\sim0.014$\,pc ($\sim2800$\,au) and an integrated flux of $12.9\pm1.3$\,mJy \citep{Barnes2023b}. The new data now begin to resolve this structure on smaller spatial scales. At $\sim10^3$\,au scales, the emission remains dominated by a single central component, hereafter C2c1a(a), with an integrated flux of $11.0\pm1.1$\,mJy (Table~\ref{tab:dendrogram_props}), comparable to the lower-resolution measurement. This indicates that the bulk of the emission remains centrally concentrated at these scales.

Only on scales of approximately 100\,au, probed by the uniformly weighted image, do we observe a significant decrease in recovered flux, to $4.9\pm0.6$\,mJy, together with a reduction in the deconvolved radius to $r_{\mathrm{deconv}}\simeq87$\,au (Table\,\ref{tab:dendrogram_props_uni}).\footnote{{The quoted $r_{\mathrm{deconv}}$ assumes that the observed compact emission is the convolution of an intrinsic Gaussian source with the Gaussian synthesised beam. We therefore subtracted the beam contribution from the two-dimensional second moment of the emission in quadrature, $r_{\mathrm{deconv}}=\left(r^2-\sigma_{\mathrm{beam}}^2\right)^{1/2}=\left(r^2-\theta_{\mathrm{beam},\mathrm{FWHM}}^2/(8\ln2)\right)^{1/2}$, where $\theta_{\mathrm{beam},\mathrm{FWHM}}=(\theta_{\mathrm{maj},\mathrm{FWHM}}\theta_{\mathrm{min},\mathrm{FWHM}})^{1/2}$ is the circularised beam full width at half maximum.}} This factor of $\sim2.2$ decrease indicates that a substantial fraction of the emission arises on spatial scales that are filtered out at the highest angular resolution. Consistent with this, the azimuthally averaged radial intensity profiles (Fig.~\ref{fig:radial_profile}) show that the source is resolved at both $\sim0.076''$ and $\sim0.033''$ resolution, while the uniformly weighted profile is narrower and recovers less total flux, again reflecting enhanced filtering of extended emission.

In addition to C2c1a(a) and C2c1b, we identify four fainter fragments within the bridging envelope, defined as C2c1a(b--e), which lie at projected separations of $\sim2400$--4300\,au. We quantified the hierarchical structure using a dendrogram analysis,\footnote{We make use of the \textsc{astrodendro} python package \citep{Rosolowsky2008a}.} adopting a minimum structure area of one beam (\texttt{min\_npix}), a minimum intensity threshold of $3\sigma$ (\texttt{min\_value}), and a minimum contrast of $1\sigma$ between structures (\texttt{min\_delta}). We used $\sigma=0.017$\,mJy\,beam$^{-1}$ for the fiducial ($\mathrm{robust}=1$) image and $\sigma=0.095$\,mJy\,beam$^{-1}$ for the uniformly weighted image.
\footnote{{The fainter sub-fragments are not detected in the uniformly weighted image because of the substantially higher noise level. The peak flux densities of the four fainter fragments in the fiducial image range from $0.068$ to $0.16$\,mJy\,beam$^{-1}$, whereas the $3\sigma$ threshold in the uniformly weighted image is 0.28\,mJy\,beam$^{-1}$.}}
The resulting dendrogram leaves are listed in Tables~\ref{tab:dendrogram_props} and \ref{tab:dendrogram_props_uni}.

\section{Mass constraints}
\label{sec:mass_constraints}

We estimated continuum masses as (see \citealp{Shirley2026})\begin{equation}
   M = \frac{S_\nu\, d^2\, R_{\mathrm{gd}}}{\kappa_\nu\, B_\nu(T_{\mathrm{dust}})},
\end{equation}
where $S_\nu$ is the integrated (primary-beam-corrected) flux density of the dendrogram leaf corresponding to the core, $d=5$\,kpc is the distance \citep{Simon2006}, $R_{\mathrm{gd}}=141$ is the total (gas plus dust)-to-(refractory-component-) dust-mass ratio (assuming a typical interstellar composition of H, He, and metals; \citealp{Draine2011})\footnote{Here we adopt a higher than typically assumed value for the dust-to-gas mass ratio, which is determined from $M_\mathrm{H}/M_\mathrm{dust}=101$ (see Table~21.3 of \citealp{Draine2011}), or $M_\mathrm{H}/M_\mathrm{dust}=101\times1.4=141$ (see Table~1.4 of \citealp{Draine2011} for $M_\mathrm{total}/M_\mathrm{H}=1.4$).}, and $\kappa_\nu=0.847$\,cm$^2$\,g$^{-1}$ is the dust opacity per gram of dust at $\nu=222.2$\,GHz, obtained by scaling $\kappa_{230\,\mathrm{GHz}}=0.899$\,cm$^2$\,g$^{-1}$ with an emissivity index $\beta=1.75$ (from \citealp{Ossenkopf1994}; see \citealp{Barnes2023b}).

Assuming optically thin emission and adopting $T_{\mathrm{dust}}=10$\,K (taken from the $2\arcsec$ NH$_3$ fits of \citealp{Wang2012}) yields masses of $M_{10\mathrm{K}} = 26.7\pm13.4$\,\msun\ and $3.6\pm1.8$\,\msun\ for C2c1a(a) and C2c1b, respectively, in the fiducial ($\mathrm{robust}=1$) image, including the adopted 50\% systematic uncertainty {(see Table~\ref{tab:dendrogram_props})}. Alternatively, assuming the core is optically thick, and adopting $T_{\mathrm{dust}}=T_{\mathrm{b,max}}$, where $T_{\mathrm{b,max}}$ is the peak continuum brightness temperature measured in each region (see Tables~\ref{tab:dendrogram_props} and \ref{tab:dendrogram_props_uni}), yields lower-limit masses (see the discussion in Sect.\,\ref{sec:discussion} on the validity of this assumption).\footnote{For the fainter fragments, $T_{\mathrm{b,max}}$ is very low and would formally imply unrealistically large masses (and densities) if interpreted as $T_{\mathrm{dust}}$, indicating that the emission is optically thin and that $T_{\mathrm{b,max}}$ does not trace the dust temperature for these components.} For C2c1a(a), we obtain $M \simeq 12.8\pm6.4$\,\msun\ in the fiducial image ($T_{\mathrm{b,max}}=16.5$\,K; Table\ref{tab:dendrogram_props}) and $M \simeq 1.9\pm0.9$\,\msun\ in the uniformly weighted image ($T_{\mathrm{b,max}}=39$\,K; Table \ref{tab:dendrogram_props_uni}). The implied volume densities on $\sim10^2$--$10^3$\,au scales are high, reaching $n_\mathrm{H2} \sim 10^9$--$10^{11}$\,cm$^{-3}$ for C2c1a(a) (Tables~\ref{tab:dendrogram_props} and \ref{tab:dendrogram_props_uni}), placing the source at the high end of the density distribution for known star-forming cores (see Fig.\,\ref{fig:lit_comparison}).

To place an approximate upper bound on the mass reservoir associated with the core, we compared it with lower-resolution (0.08\,pc) Band~3 fluxes from \citet{Barnes2021c}, for which optical-depth effects are expected to be less severe. Using the integrated Band~3 flux density of $S_\nu=2.38$\,mJy {(measured within an aperture of 2.9\arcsec, or 14\,500\,au)} at 90.664\,GHz, we derive $M\simeq113\pm57$\,\msun\ for $T_{\mathrm{dust}}=10$\,K and $M\simeq24.7\pm12.4$\,\msun\ for $T_{\mathrm{dust}}=T_{\mathrm{b,max}}=38.7$\,K, including the adopted 50\% systematic uncertainty. These estimates bracket a plausible mass range of $\sim25$--$110$\,\msun, but we expect that the true value is likely closer to the lower-temperature (higher mass) limit on these larger scales {(which are closely matched in scale to the NH$_3$ fits of \citealp{Wang2012})}. Although the Band~3 measurements include emission from the surrounding envelope, C2c1a(a) dominates the continuum flux across all spatial scales probed to date, suggesting that a substantial fraction of this mass reservoir is physically associated with the compact core. {The present data do not, however, determine how much of this reservoir is dynamically bound to the compact component or how much may accrete onto it from the surrounding clump.}

\section{Gas kinematics, chemistry, and outflows}
\label{sec:chemistry}

\begin{figure}[!t]
\centering
\includegraphics[trim={0 20 0 0}, clip, width=\columnwidth]{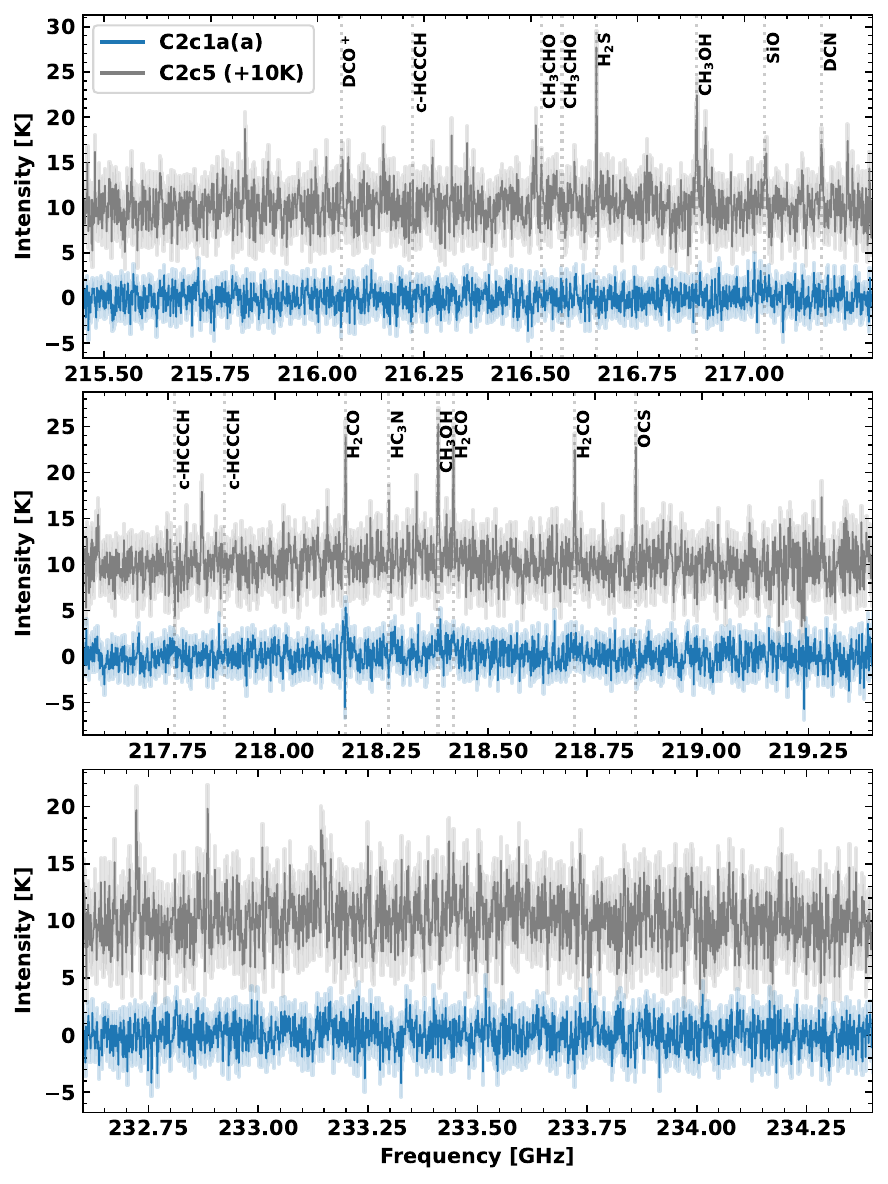} \\
\includegraphics[trim={0 0 0 5}, clip, width=\columnwidth]{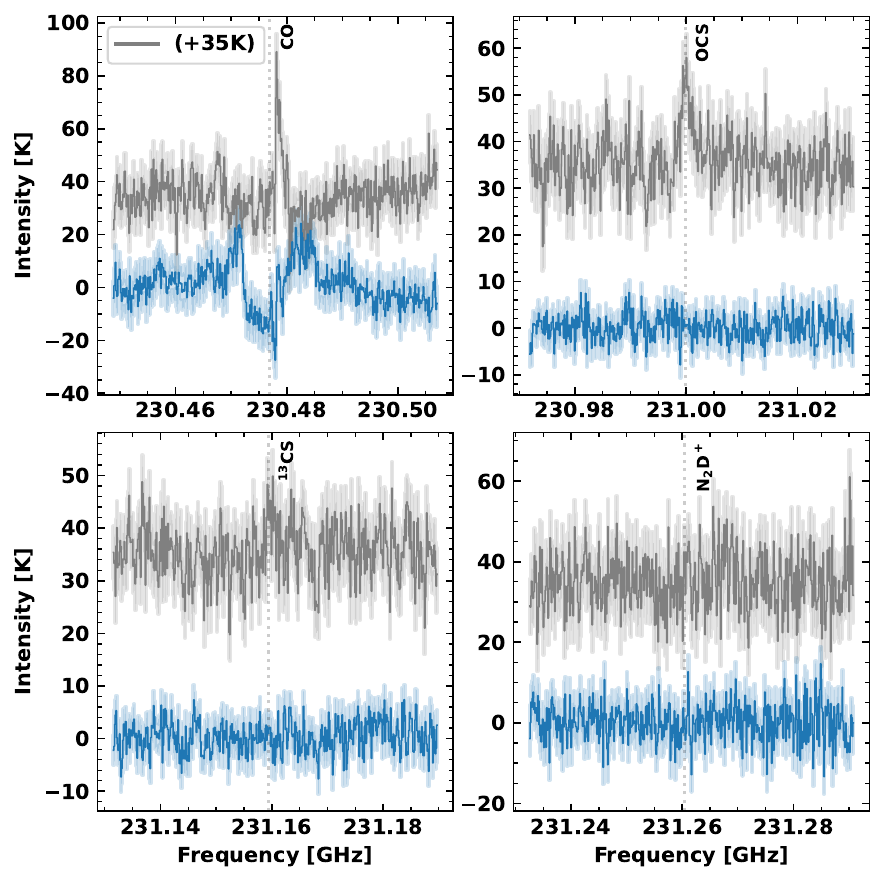}
\caption{Aperture-averaged spectra extracted within a circular aperture of $0.08\arcsec$ towards C2c1a(a) (blue) and the nearby, more actively star-forming core C2c5 (grey; centred at $\alpha_{\rm ICRS} = 280.710746^\circ$, $\delta_{\rm ICRS} = -4.054494^\circ$).\ They are offset by $+10$\,K in the upper panels and $+35$\,K in the lower panels. Vertical dotted lines mark the systemic rest frequencies of identified transitions. 
}
\label{fig:aperture_spectra}
\end{figure}

\begin{figure*}
\centering
\includegraphics[width=\textwidth]{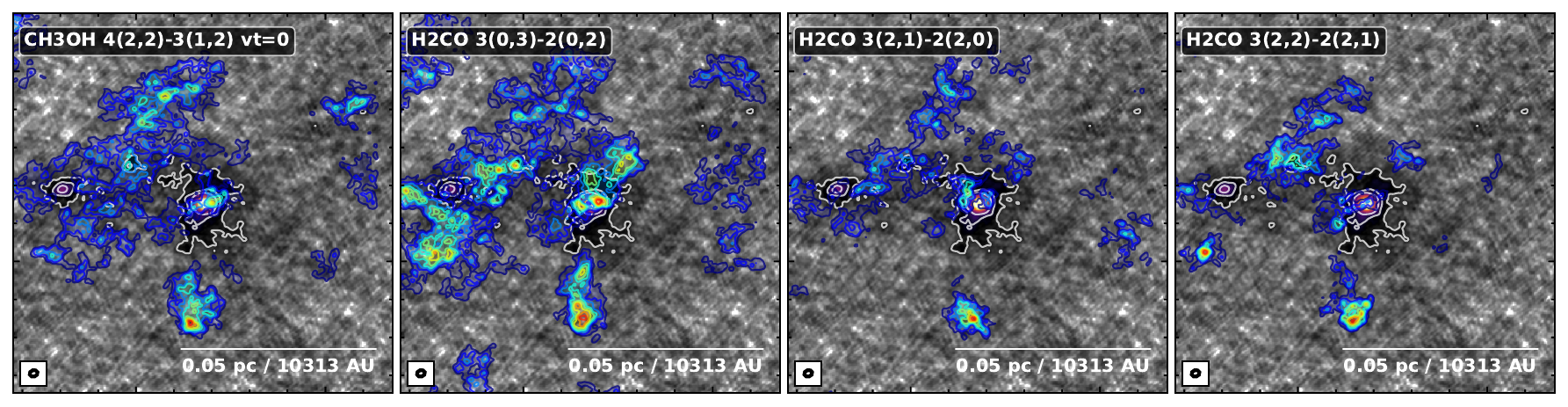}
\caption{
   Integrated intensity maps of molecular transitions
    towards the C2c1a region (see Fig.\,\ref{fig:moment_maps_zoom} for a zoomed-in version of this figure).
    \textit{From left to right}:
    CH$_3$OH\,(4$_{2,2}$--3$_{1,2}$)\,$v_{\rm t}=0$ and three H$_2$CO transitions:
    3$_{0,3}$--2$_{0,2}$,
    3$_{2,1}$--2$_{2,0}$,
    and 3$_{2,2}$--2$_{2,1}$.
    The underlying grey-scale image in each panel is the 1\,mm continuum emission
    (Briggs robust $+1$ weighting), displayed with a square-root stretch.
    White contours trace the continuum emission at
    $2$, $5$, $10$, and $100\sigma$, where $\sigma$ is the continuum rms.
    Coloured contours (with semi-transparent filled levels) show the integrated
    line emission in steps of 0.1 times the peak moment~0 intensity in each panel.
    The colour scale and spatial coverage are identical to those adopted for the
    middle panel of Fig.~\ref{fig:continuum_maps}, enabling a direct comparison
    between the continuum morphology and the molecular line distribution.
    Each panel is centred at
    $\alpha_{\mathrm{J2000}} = 18^{\mathrm h}42^{\mathrm m}49.819^{\mathrm s}$,
    $\delta_{\mathrm{J2000}} = -04^{\circ}03'25.39''$
    and spans $4'' \times 4''$.
    The synthesised beam is shown in the lower-left corner of each panel.
    A scale bar corresponding to $0.05$\,pc ($\sim1.0\times10^{4}$\,au) at a
    distance of 5\,kpc is indicated.
}
\label{fig:moment_maps}
\end{figure*}

\begin{figure}[t]
\centering
\includegraphics[width=\columnwidth]{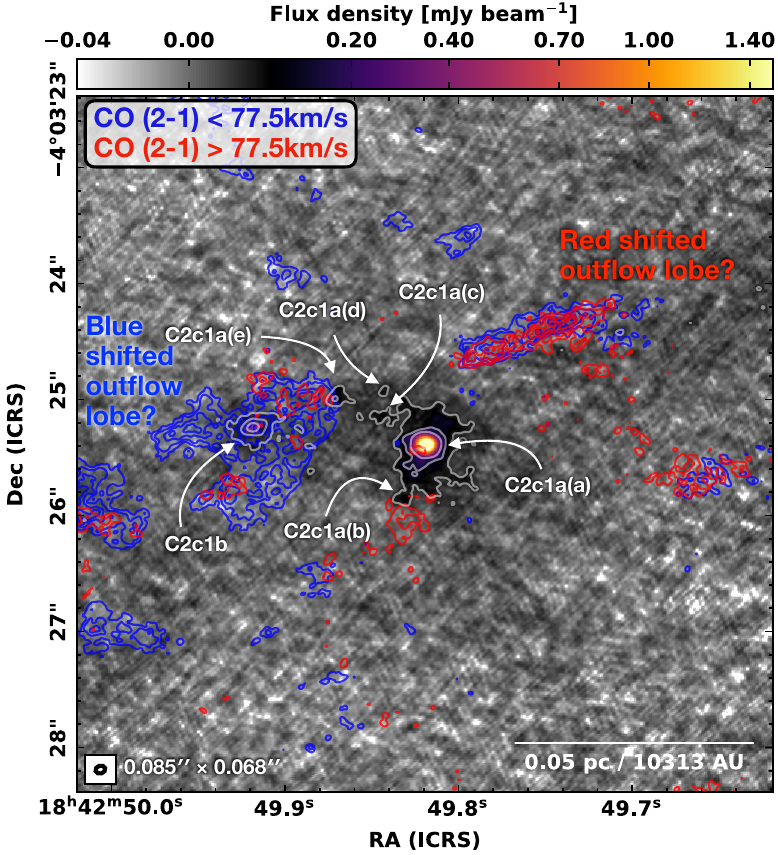}
\caption{ALMA 1\,mm continuum emission towards the C2c1 region (see also Fig.~\ref{fig:continuum_maps}).
White contours trace the continuum at $[2,5,10,100]\sigma$, where $\sigma$ is the rms measured from the continuum map.
Contours show the peak intensity map of the CO cube, overlaid at $[5,6,7,8,9,10]\times2$~mJy~beam$^{-1}$ below 77.5\,\kms in blue and above 77.5\,\kms in red.
}
\label{fig:cont_briggs_co}
\end{figure}

We show the 0.080\arcsec\ aperture-averaged spectra towards C2c1a(a) in Fig.~\ref{fig:aperture_spectra}, together with spectra extracted towards the nearby actively star-forming core C2c5 for comparison (see \citealp{Zhang2015, Barnes2023b} for similar comparisons). {While C2c5 exhibits multiple bright, high-S/N molecular transitions across the observed spectral windows, the spectral setup reveals only limited molecular-line emission towards C2c1a(a).} At the compact continuum peak we do not detect strong hot-core tracers such as OCS, CS isotopologues, bright CH$_3$OH transitions, or SiO at $>3\sigma$ significance. Weak H$_2$CO and CO emission is present at the position of C2c1a(a); however, the line profiles show prominent negative features, suggesting that part of the emission arises from extended envelope material that is either partially filtered by the interferometric observations or absorbed by optically thick dust.

To investigate the origin of the weak H$_2$CO (and CH$_3$OH) emission, we present integrated-intensity maps of selected transitions over the full velocity range of the source in Fig.~\ref{fig:moment_maps}.\footnote{The integrated-intensity maps are constructed from velocity sub-cubes extracted over 40 to 110\,km\,s$^{-1}$ and converted to brightness temperature units. The rms noise is estimated iteratively from emission-free regions, and an expansion mask is generated using high- and low-significance thresholds of 5$\sigma$ and 3$\sigma$, respectively. This mask is applied to the cube to isolate spatially and spectrally coherent emission, and moment maps are then computed from the masked data, thereby minimising noise contributions while preserving extended low-level structure.} These maps reveal extended structures surrounding the compact continuum peak that do not closely follow the continuum morphology. A local maximum in several transitions lies near C2c1a(a), but it is offset by approximately one synthesised beam from the continuum maximum (see also Fig.\,\ref{fig:moment_maps_zoom}). The extended morphology, together with the lack of direct spatial correspondence with the compact continuum structure, suggests that a substantial fraction of the line emission originates in the surrounding envelope and may also be affected by optically thick dust. However, the spatial coincidence of the emission peak with C2c1a(a) indicates that a compact contribution cannot be ruled out.

To assess whether these weak detections constrain the thermal conditions, we performed simple RADEX excitation tests \citep{vanderTak2007}. For representative dense-gas parameters consistent with the continuum constraints ($n_{\mathrm{H_2}}\sim10^{8}$\,cm$^{-3}$, $\Delta v\sim1$--1.5\,km\,s$^{-1}$, and $N(\mathrm{H_2CO})\sim10^{14}$\,cm$^{-2}$), a kinetic temperature of $\sim40$\,K (similar to $T_{\mathrm{b,max}}$) predicts H$_2$CO line brightness exceeding the $\sim3\sigma$ detection threshold of $\sim6$\,K in our setup (e.g. for the $3_{0,3}$--$2_{0,2}$ transition; a higher assumed H$_2$CO abundance would further increase the predicted emission).\footnote{We attempted a rotational-diagram analysis of the detected H$_2$CO transitions to constrain the excitation conditions. However, the three available lines do not define a physically meaningful solution, yielding a non-negative slope and thus no valid rotational temperature. In particular, the lowest-excitation transition ($3_{0,3}$--$2_{0,2}$; $E_u/k\sim21$\,K) shows pronounced absorption near the systemic velocity, while the higher-excitation lines ($E_u/k\sim68$\,K) remain primarily in emission (see also Fig.\,\ref{fig:moment_maps_zoom}). This behaviour is inconsistent with a single-component, optically thin excitation scenario and likely reflects a combination of foreground absorption, optical depth effects, and spatial filtering of extended emission. As a result, the line ratios do not trace a common gas component, and a simple rotational-diagram interpretation is not applicable.} Under these assumptions, emission originating from the compact core would be expected to peak at the continuum maximum and be readily detected. The absence of such emission therefore suggests that the excitation conditions within the compact component differ from these simple expectations, for example due to strong dust optical depth, suppressed gas-phase abundances (based on our crude assumptions for this modelling), or non-local thermodynamic equilibrium effects that reduce the emergent line brightness.

We also show detectable CO emission within the C2c1 field in Fig.~\ref{fig:cont_briggs_co}. The brightest CO structure is spatially offset from C2c1a(a) and lies closer to C2c1b (see \citealp{Barnes2023b}). However, as also noted by \citet{Barnes2023b}, the red- and blue-shifted components are not cleanly separated in velocity space, both lying close to $\sim77.5$\,km\,s$^{-1}$, and partially overlap spatially in the northern lobe. Moreover, this velocity is offset from the systemic velocities of both C2c1a(a) ($79.4$\,km\,s$^{-1}$) and C2c1b ($79.1$\,km\,s$^{-1}$). \citet{Wang2011} estimated that C2c1 has the lowest mass outflow rate in the region, $0.73\times10^{-5}$,M$_\odot$,yr$^{-1}$, while \citet{Liu2021} found no associated SiO outflow. Consistent with this, we do not detect SiO emission in the integrated intensity maps of the region, while clear outflow morphologies are observed towards more actively star-forming cores such as C2c5, indicating that such emission should be detectable if present. Given these ambiguities, even in the present data we cannot definitively attribute the observed CO emission to an outflow or associate it with any compact continuum peak (see Sect.~\ref{sec:discussion} for further discussion).

\section{Discussion}
\label{sec:discussion}

{The new ALMA Band~6 observations place C2c1a(a) among the most compact high-mass core candidates currently resolved on scales of a few hundred au. At $\sim175$--400\,au resolution, the source remains a single continuum peak, with no evidence of multiplicity within the central $\sim175$\,au. Figure~\ref{fig:lit_comparison} traces how this compact structure emerges from the fragmentation hierarchy of Cloud~C. At larger scales, C2c1 contains the two previously identified cores C2c1a and C2c1b, separated by $\sim7500$\,au \citep{Barnes2021c,Barnes2023b}; across the same region, the fiducial Band~6 image identifies six components, comprising the dominant C2c1a(a) peak, C2c1b, and four fainter leaves in the intervening envelope. The four leaves lie at projected separations of $\sim2400$--4300\,au from C2c1a(a), whereas no additional component is detected within the central $\sim175$\,au. The fragmentation is therefore hierarchical but strongly concentrated, with increased resolution revealing additional structure across the envelope rather than a succession of fragments towards the central peak.}

{This behaviour provides a direct point of comparison with other high-mass regions, many of which fragment on larger scales (e.g. \citealp{Molinari2025,Sanchez-Monge2025,Motte2022,Ginsburg2022b,Sanhueza2019,Morii2024}). Studies at spatial resolutions of $\sim160$--1000\,au often resolve massive cores into additional compact components (e.g. \citealp{Brogan2016,Olguin2022,Budaiev2024,Yoo2025,Luo2026}), although the measured fragmentation can remain limited or strongly unequal where a direct parent-to-fragment comparison is possible. For example, only about one-third of the lower-resolution Sgr~B2 sources fragment, producing two components on average \citep{Budaiev2024}, while an equal division of flux among the fragments in W51 is uncommon \citep{Yoo2025}. In this qualitative comparison, C2c1a(a) resembles the low-fragmentation cases: the surrounding envelope contains several components, but the central branch remains dominated by one compact peak.}

{Expressed relative to the previously inferred Jeans scale ($\lambda_{\mathrm{J}}\sim700$\,au; \citealp{Barnes2023b}), the new data detect no additional fragment within $\lesssim0.2\,\lambda_{\mathrm{J}}$, while the fainter envelope leaves lie at projected separations of a few Jeans lengths (noting the caveats in interpreting projected separations; \citealp{Barnes2026}). This scale dependence suggests that fragmentation occurs predominantly within the larger envelope, while the central $\sim10^2$\,au component remains highly concentrated. In the uniform weighted (robust $-2$) continuum image, no additional compact fragments are detected above a $5\sigma$ threshold within the central $\sim175$\,au structure, corresponding to an approximate unresolved mass sensitivity of $\sim0.2$--1.1\,\msun\ for dust temperatures of 40--10\,K. At the same time, C2c1a(a) is resolved and shows clear evidence of extended emission that is increasingly filtered at higher angular resolution. The continuum morphology therefore describes a centrally peaked structure embedded within a broader reservoir rather than a completely isolated compact object (see Fig.~\ref{fig:radial_profile}).}

{Having established that a compact central concentration persists to these scales, the principal remaining uncertainty is its mass and physical nature (see \citealp{Barnes2023b}).} If the dust remains cold ($\sim10$\,K), the compact core can still contain several tens of solar masses on sub-1000\,au scales. Conversely, if the high continuum brightness temperature reflects internal heating ($T_{\mathrm{dust}}\sim40$\,K), the few-\msun\ mass inferred under the optically thin assumption should be regarded as a conservative lower limit, since optically thick dust would both elevate the observed brightness temperature and obscure a fraction of the continuum emission. {Independent of the conditions within the innermost $\sim100$\,au, the lower-resolution Band~3 data indicate that the compact core is embedded within a much larger reservoir with a plausible mass of $\sim25$--110\,\msun\ \citep{Barnes2021c,Barnes2023b}. C2c1a(a) dominates the continuum emission across the spatial scales probed to date, suggesting that a substantial fraction of this reservoir is physically associated with the compact core (see the hierarchical structure shown in Fig.~\ref{fig:lit_comparison}). The data do not, however, determine how much of this material is bound to the compact component or may subsequently accrete onto it. The continuum measurements therefore support a stratified structure in which a compact, potentially warm component is embedded within a colder and more massive envelope.}

{The molecular-line and outflow diagnostics provide complementary constraints on the evolutionary state of C2c1a(a). Only limited molecular-line emission is detected towards the continuum peak, while the weak H$_2$CO and CO emission is largely extended and offset from the continuum maximum (Fig.~\ref{fig:moment_maps}). This morphology suggests that the detected lines primarily trace the surrounding gas rather than the compact peak itself. The absence of centrally concentrated line emission may indicate a low temperature, although continuum or line optical depth could also suppress the line contrast at the core position \citep[see][]{DeSimone2020}. We likewise find no outflow that can be robustly associated with C2c1a(a). CO emission is present nearby, but the brightest structures are spatially and kinematically offset from the core and lie closer to C2c1b (Fig.~\ref{fig:cont_briggs_co}; see also \citealp{Barnes2023b}). A weak, deeply embedded, or unfavourably oriented outflow cannot be excluded, but the present data reveal no clear outflow signature from the compact source.}

{The luminosity-to-mass ratio of the parent clump provides an independent indication of its early evolutionary state. Cloud~C has $L/M\simeq0.5\,L_\odot\,M_\odot^{-1}$, while empirical evolutionary schemes generally associate $L/M\lesssim1\,L_\odot\,M_\odot^{-1}$ with the earliest phases of high-mass star formation \citep[e.g.][]{Molinari2008,Molinari2016,Elia2010,Elia2017,Elia2021,Traficante2023}. The measured value lies within the approximate $L/M\sim0.1$--$1\,L_\odot\,M_\odot^{-1}$ range occupied by the 70\,$\mu$m-dark clumps in the `ALMA Survey of 70\,$\mu$m dark High-mass clumps in Early Stages' (ASHES) sample \citep{Sanhueza2019,Sabatini2022,Morii2023}, and near the low end of the much broader $L/M\sim0.05$--$450\,L_\odot\,M_\odot^{-1}$ range covered by ALMAGAL \citep{Molinari2025,Coletta2025}. By comparison, the young and evolved protoclusters in ALMA-IMF have mean clump-scale ratios of approximately 20 and $76\,L_\odot\,M_\odot^{-1}$, respectively \citep{Dell'Ova2024}. Although these surveys adopt different luminosity measurements, mass estimates, and physical apertures, they consistently place Cloud~C among the lowest-$L/M$ high-mass clumps. In addition, it is worth highlighting that the \textit{Herschel} measurement of L/M does not resolve C2c1a(a) from nearby, more evolved sources, so part of the measured luminosity may arise outside the compact core. The quoted ratio is therefore a clump-scale constraint rather than a direct measurement of $L/M$ for C2c1a(a), and the heating attributable to the compact source is likely bounded by this value. Taken together, the limited molecular-line emission, absence of a robustly associated outflow, and low $L/M$ of the parent clump remain consistent with C2c1a(a) being at a very early evolutionary stage, even if it already contains a weak embedded heating source.}

{Figure~\ref{fig:lit_comparison} shows that the allowed mass range of C2c1a(a) overlaps that of several candidate young high-mass cores, including the ALMA-IMF pre-stellar-core candidates \citep{Valeille-Manet2025}, G11.11$-$P6$-$SMA1 \citep{Wang2014}, CygX$-$N53$-$MM2 \citep{Duarte-Cabral2013}, and G28.34 C1-S \citep{Tan2013,Tan2016,Kong2018}. C2c1a(a), however, extends this comparison to much smaller scales: its deconvolved radius of only $\sim87$\,au places it at the compact edge of the candidate population, and no further fragment is detected within the central $\sim175$\,au despite a mass sensitivity of $\sim0.1$--0.5\,\msun. The surrounding envelope instead fragments at separations of a few thousand au. The uncertain dust temperature leaves the nature of the compact emission ambiguous: it may trace either a cold, genuinely massive core or a warmer, lower-mass component embedded within the surrounding reservoir. Nevertheless, this combination of a single central concentration on $\sim100$\,au scales, a potentially substantial envelope, and no robust line or outflow signature distinguishes C2c1a(a) from most objects in the comparison and makes it an unusually stringent test of a young high-mass core candidate.}

\section{Conclusions}
\label{sec:conclusions}

We present high-angular-resolution ALMA Band~6 observations of the dragon egg, or C2c1a, in the dragon cloud (G028.37+00.07 or Cloud~C), obtained as a stringent follow-up test of its classification as {a compact high-mass pre-stellar core candidate} \citep{Barnes2023b}. These data provide a detailed view of a compact core candidate in an infrared dark cloud. Our main findings are as follows:

\begin{itemize}
\item C2c1a(a) remains a single, compact continuum peak down to $\sim175$\,au, with no evidence of multiplicity. The source is resolved, centrally concentrated, and among the most compact high-mass core candidates studied at comparable scales, with fragmentation confined to larger ($\sim10^3$\,au) separations within the surrounding envelope.

\item Only a few molecular transitions are detected towards the compact continuum peak given the observed spectral setup and sensitivity. Strong hot-core tracers are not detected, and the weak molecular emission present is predominantly extended and offset from the continuum maximum. This indicates that the available line tracers do not directly probe the compact inner structure, potentially due to a combination of low excitation, chemical depletion, or continuum optical depth effects.

\item While CO emission is detected in the region, no high-velocity structure can be robustly associated with C2c1a(a). The absence of a clearly associated outflow, together with the lack of centrally peaked line emission, is consistent with a very early evolutionary stage, although a weak or deeply embedded outflow cannot be excluded.

\item Continuum-based mass estimates for the compact component span $\sim1.9$--$12.8$\,\msun\ when adopting $T_{\mathrm{dust}}=T_{\mathrm{b,max}}$, while colder assumptions permit substantially higher masses ($\sim$\,30\,\,\msun\ at $T_{\mathrm{dust}}=10$\,K). Comparison with lower-frequency data indicates the presence of a larger-scale mass reservoir of $\sim100$\,\msun.
\end{itemize}

Taken together, these observations show that C2c1a(a) is a compact concentration embedded within a larger and potentially massive cold reservoir. The source remains dominated by a single compact continuum peak at the current resolution and sensitivity limits, with few detected molecular lines and no robustly associated outflow. Even if the high continuum brightness temperature reflects heating by an embedded protostar, these properties indicate that C2c1a(a) remains at a very early evolutionary stage. The presence of such a compact and potentially massive structure before the emergence of prominent protostellar activity would be consistent with the initial conditions invoked by monolithic core-accretion models. However, the current observations do not uniquely distinguish this scenario from other possible evolutionary pathways, including those in which the source continues to gain mass through accretion from the surrounding reservoir. C2c1a(a) may therefore represent a very young compact high-mass star-forming region, providing a rare view of the earliest stages of collapse and mass assembly within a still-dominant cold envelope.

\begin{acknowledgements}
We are grateful to the referee for their constructive suggestions, and to the ALMA-IMF team for valuable discussions during the revision of this manuscript.
A.S-M. acknowledges support from the PID2023-146675NB grant funded by MCIN/AEI/10.13039/501100011033, and by the programme Unidad de Excelencia Mar\'{\i}a de Maeztu CEX2020-001058-M, as well as support from the RyC2021-032892-I grant funded by MCIN/AEI/10.13039/501100011033 and by the European Union `Next GenerationEU'/PRTR.
I.J-.S acknowledges funding from grant PID2022-136814NB-I00 funded by the Spanish Ministry of Science, Innovation and Universities/State Agency of Research MICIU/AEI/ 10.13039/501100011033 and by “ERDF/EU”.
\end{acknowledgements}

\bibliographystyle{aa}
\bibliography{references}

\begin{appendix}
\nolinenumbers

\section{{Literature mass--radius comparison}}
\label{sec:app_literature_comparison}

{Figure~\ref{fig:lit_comparison} shows the fragmentation hierarchy of the C2C1 region. For the `ALMA InfraRed-Dark Cloud' (ALMA-IRDC) sample \citep{Barnes2021c}, the markers show masses calculated using the \textit{Herschel}-derived temperatures, while the vertical ranges extend from the background-subtracted mass minus 50\% to the total mass plus 50\%. For C2c1a and C2c1b, the markers show the published total masses, with lower bounds set by the background-subtracted masses and upper bounds set by the uncertainties on the total masses \citep{Barnes2023b}. For both measurements of C2c1a(a), the markers are placed at the midpoint of the displayed mass range. Here, the lower bounds correspond to the masses calculated using the peak brightness temperatures minus their uncertainties, while the upper bounds correspond to the 10\,K masses plus their uncertainties. The other components identified in this work are shown at their 10\,K masses with 50\% uncertainties (see Tables~\ref{tab:dendrogram_props} and~\ref{tab:dendrogram_props_uni}).}

{To place the C2C1 hierarchy in a broader context, we assembled a comparison sample of 1988 selected measurements from 21 studies \citep{Bontemps2010,SanchezMonge2013,Henshaw2016a,Brogan2016,Fontani2016,Kong2016,Fontani2018,Beuther2018,Liu2018,Kong2018,Wang2014,Duarte-Cabral2013,Peretto2013,Beltran2021,Olguin2022,Morii2023,Budaiev2024,Li2024b,Valeille-Manet2025,Yoo2025,Luo2026}. We retained sources with finite, positive masses and radii within $30\leq R\leq30{,}000$\,au and $0.01\leq M\leq300\,M_\odot$, while preserving the published upper and lower limits. We did not rescale the masses to common dust opacities, temperatures, distances, gas-to-dust ratios, or background treatments, as these choices contribute to the systematic uncertainty of the comparison. Where available, we adopted the published physical or equivalent radius. For catalogues reporting a deconvolved diameter or two Gaussian full widths at half maximum (FWHMs) due to elliptical constraints, we calculated the radius as $R=D/2$ or $R=\sqrt{\mathrm{FWHM}_{\mathrm{maj}}\mathrm{FWHM}_{\mathrm{min}}}/2$, respectively.}

{The largest samples provide broad context across different physical scales. For the ASHES catalogue, we used all 839 cores from \citet{Morii2023}, adopting the published half-geometric-mean deconvolved FWHM radius and optically thin 1.3 mm dust mass. The 118 Northern Extended Millimetre Array (NOEMA) CORE components from \citet{Beuther2018} use the published equivalent area radius and optically thin dust masses calculated with a gas-to-dust ratio of 150, $\kappa_{220\,\mathrm{GHz}}=0.9$\,cm$^2$\,g$^{-1}$ per gram of dust, and one H$_2$CO temperature per region. These masses are lower limits because of spatial filtering (see \citealp{Beuther2018}). For the 107 infrared dark cloud cores from \citet{Liu2018}, we retained the published equivalent area radius and flux-recovery-corrected 1.3 mm dust mass, which assumes $T_{\mathrm{dust}}=20$\,K, a gas-to-dust ratio of 141, and $\kappa_{\mathrm{dust}}=0.899$\,cm$^2$\,g$^{-1}$. The 82 ALMA-IMF cores from \citet{Valeille-Manet2025} comprise 30 pre-stellar candidates and 52 protostellar cores. We converted the published deconvolved FWHM to a radius by dividing it by two and retained the published temperature-bracketed or PPMAP-temperature masses.}

{At smaller physical scales, we selected 122 high-confidence, resolved, core-classified sources in Sgr~B2 from \citet{Budaiev2024}. We calculated their radii from the 3 mm Gaussian fits as $R=\sqrt{\mathrm{FWHM}_{\mathrm{maj}}\mathrm{FWHM}_{\mathrm{min}}}/2$ and used the published 3 mm mass lower limits, which assume $T_{\mathrm{dust}}=50$\,K. Sources classified as H\,{\sc ii} regions or candidates were excluded. For W51, we selected the 55 resolved dust-dominated pre/protostellar objects from \citet{Yoo2025}. Their plotted radius is the published semi-major axis radius containing 95\% of a Gaussian, $R_{95}=\mathrm{FWHM}_{\mathrm{maj}}/\sqrt{2\ln2}$, and their 3 mm masses are lower limits calculated at 40, 100, or 300\,K according to the source classification. Objects flagged as potentially free--free dominated were excluded. The Digging into the Interior of Hot Cores with ALMA (DIHCA) survey catalogue of \citet{Luo2026} provides half-FWHM radii but not condensation masses. We therefore calculated masses for 439 objects inside the radius boundary from the published 1.33 mm fluxes, region distances, and temperatures, following the authors' core-mass conversion with a gas-to-dust ratio of 100, $\kappa_{1.33\,\mathrm{mm}}=0.899$\,cm$^2$\,g$^{-1}$ per gram of dust, and a frequency of 226\,GHz. These derived masses are shown only as a low-weight contextual distribution because compact-source temperatures, optical depth, free--free emission, and spatial filtering introduce substantial systematic uncertainties.}

{The intermediate-scale comparison includes 72 ammonia cores from \citet{SanchezMonge2013}. We divided the published deconvolved FWHM diameter by two and retained the NH$_3$-abundance-derived masses and the published quiescent starless, perturbed starless, protostellar, and possible protostellar classifications. Four secondary velocity components without independent sizes and one core without a mass were excluded. For the 13 G035.39$-$00.33 H6 cores from \citet{Henshaw2016a}, we used the published equivalent area radius and full leaf mass from their dendrogram analysis. The background-subtracted values represent an alternative boundary treatment and were retained as uncertainties rather than separate sources. The 68 young-clump fragments from \citet{Fontani2016,Fontani2018} use $R=D/2$, where $D$ is the published equivalent diameter of the area within the source polygon, together with the published optically thin dust masses. These masses assume region-average temperatures, a gas-to-dust ratio of 100, and an opacity normalised to $\kappa_{230\,\mathrm{GHz}}=0.899$\,cm$^2$\,g$^{-1}$. We treated the masses of fragments associated with 24-$\mu$m emission as upper limits where the adopted clump temperature may underestimate the source temperature.}

{For Cygnus~X, we used 22 compact fragments from \citet{Bontemps2010}, pairing the native interferometric 1.3 mm mass with half the measured deconvolved FWHM. We preserved the unresolved size of N3-MM3 as an upper limit, omitted the diffuse N40-MM1 detection from the selected sample, and excluded N53-MM2 because the later measurement from \citet{Duarte-Cabral2013} is plotted separately. For N53-MM2, we adopted the published spectral-energy-distribution envelope mass and half of the 4000 au FWHM aperture. For SDC335, only MM2 from \citet{Peretto2013} lies inside the mass boundary; we used half its published deconvolved diameter and its 3.2 mm dust mass. MM1 is not plotted because its central mass estimate is $545\,M_\odot$.}

{Additionally, we included the published physical radii and temperature-dependent gas masses of 14 NGC~6334I components \citep{Brogan2016}; the parent core and four G335.579$-$0.272 fragments, with radii derived from the deconvolved Gaussian FWHM axes and two optically thick mass lower limits \citep{Olguin2022}; four G31.41$+$0.31 hot-core fragments with published Gaussian-fit radii and 3.5 mm masses \citep{Beltran2021}; and 20 G333.23$-$0.06 multiple-system members with published radii and ambient gas-mass lower limits \citep{Li2024b}. For G11.11$-$P6$-$SMA1, we derived the radius from the published deconvolved Gaussian axes and retained the 11-K dust mass \citep{Wang2014}. For G28.34, we included C1-N and four measurements associated with the C1-S region. The published radius and temperature-bracketed mass of C1-N are from \citet{Tan2013}, who first identified both C1-N and C1-S. \citet{Tan2016} subsequently identified the nearby protostellar sources C1-Sa and C1-Sb, while their published radii and temperature-bracketed masses, together with those of the C1-S inner and outer apertures, are from \citet{Kong2018}. The C1-S inner and outer values are nested aperture measurements rather than independent fragments. W43-MM1 cores 3 and 6 from \citet{Nony2018} were not added separately because both are represented in the newer ALMA-IMF catalogue of \citet{Valeille-Manet2025}. Similarly, the ASHES pilot catalogue of \citet{Sanhueza2019} was not added because its sources are included in the full \citet{Morii2023} catalogue.}

{The complete compilation will be made available online. The release contains 2444 provenance records, including measurements that are outside the displayed range, superseded duplicate measurements, and primary-table entries that lack one plotted quantity. The Boolean \texttt{selected} column identifies the 1988 records used in Fig.~\ref{fig:lit_comparison}.}

\section{{Literature sensitivity--resolution comparison}}
\label{sec:app_sensitivity_resolution}

\begin{figure}[!t]
\centering
\includegraphics[width=\columnwidth]{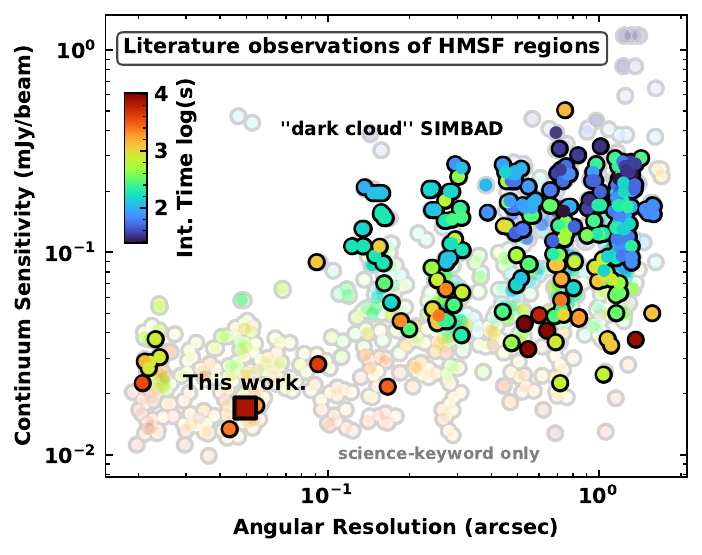}
\caption{Continuum sensitivity versus angular resolution for ALMA Band~6 (12 m-only) observations retrieved from the ALMA Science Archive using the science-keyword queries `high-mass star formation' and `pre-stellar cores, IRDCs' in August 2026 (faded points). Points are coloured by the logarithm of the on-source integration time (see the colour bar in the upper left). Targets shown in solid colours and outlined in black are those classified as `dark cloud', for which at least one SIMBAD source with this object type falls within the ALMA footprint according to the archive's automatic object-type association. This classification is incomplete and should therefore be regarded as representative only. The square marks the Cloud~C C2c1 observations presented in this work (see Sect.~\ref{sec:obs}).
    }
\label{fig:lit_comparison_alma}
\end{figure}

{To place the angular resolution and continuum sensitivity of the C2c1 observations in context, we queried the ALMA Science Archive in August 2026 for Band~6 observations obtained with the 12 m array and associated with either of the science keywords `high-mass star formation' or `pre-stellar cores, IRDCs'. Each point in Fig.~\ref{fig:lit_comparison_alma} represents an individual archive entry rather than a unique target or project. We used the archive-reported angular resolution, continuum sensitivity, and on-source integration time without rescaling for differences in bandwidth, spectral setup, visibility weighting, mosaicking strategy, or observing conditions. The comparison is therefore intended to show the range of available observations, rather than to provide a homogeneous sensitivity assessment.}

{The C2c1 observations lie towards the high-resolution, high-sensitivity part of the archive distribution, including among the subset associated with dark clouds. This supports their use as a stringent test of compact structure in a young high-mass core candidate, but does not imply that the archive sample is complete or uniformly selected. In particular, the archive values describe the observation entry and do not correspond exactly to either of the visibility-weighting choices used for the images in this work (Sect.~\ref{sec:obs}).}

\section{Spectral setup}
\label{sec:app_spectral_setup}

Table~\ref{tab:spectral_setup} summarises the spectral configuration of the ALMA Band~6 observations. Three wideband spectral windows were configured to maximise continuum sensitivity, each providing a bandwidth of 1.875\,GHz with coarse spectral resolution. In addition, four narrowband spectral windows were placed on selected molecular transitions, including N$_2$D$^+$, $^{13}$CS, OCS, and CO, to probe the dense gas, chemistry, and kinematics of the region at higher spectral resolution. The narrowband setup provides a velocity resolution of $\sim0.16$ to $0.18$\,km\,s$^{-1}$ over a total velocity coverage of $\sim76$\,km\,s$^{-1}$, enabling detailed analyses of the line profiles and gas dynamics.

\begin{table*}
\centering
\small
\caption{Spectral setup for the ALMA Band~6 observations. }
\label{tab:spectral_setup}
\begin{tabular}{ccccccc}
\hline\hline
Freq.$^{a}$ & SPW$^{b}$ & Ch.$^{c}$ & BW$^{d}$ & Res.$^{e}$ & Vel. BW$^{f}$ & Vel. Res.$^{g}$ \\
(GHz) &  &  & (MHz) & (MHz) & (km\,s$^{-1}$) & (km\,s$^{-1}$) \\
\hline
    216.4445 & cont. & 3840 & 1875 & 1.129 & 2597.6 & 1.564 \\
    218.5450 & cont. & 3840 & 1875 & 1.129 & 2572.6 & 1.549 \\
    233.5480 & cont. & 3840 & 1875 & 1.129 & 2407.3 & 1.449 \\
    231.3218 & N$_2$D$^+$ & 960 & 58.6 & 0.122 & 76.0 & 0.158 \\
    231.2207 & $^{13}$CS & 960 & 58.6 & 0.141 & 76.0 & 0.183 \\
    231.0610 & OCS       & 960 & 58.6 & 0.141 & 76.0 & 0.183 \\
    230.5380 & CO        & 960 & 58.6 & 0.122 & 76.2 & 0.159 \\
\hline
\end{tabular}
\tablefoot{The wideband spectral windows are configured for continuum sensitivity, while the narrowband spectral windows target selected molecular transitions at high spectral resolution.
\tablefoottext{a}{Central rest frequency of each spectral window.}
\tablefoottext{b}{Spectral window designation; `cont.' denotes wideband continuum windows.}
\tablefoottext{c}{Number of effective spectral channels per window.}
\tablefoottext{d}{Total bandwidth of the spectral window.}
\tablefoottext{e}{Native channel width in frequency units.}
\tablefoottext{f}{Total velocity coverage corresponding to the listed bandwidth at the rest frequency.}
\tablefoottext{g}{Velocity resolution corresponding to the native channel width.}
}
\end{table*}

\section{Zoomed moment maps}
\label{sec:app_moment_maps}

\begin{figure*}
\centering
\includegraphics[width=\textwidth]{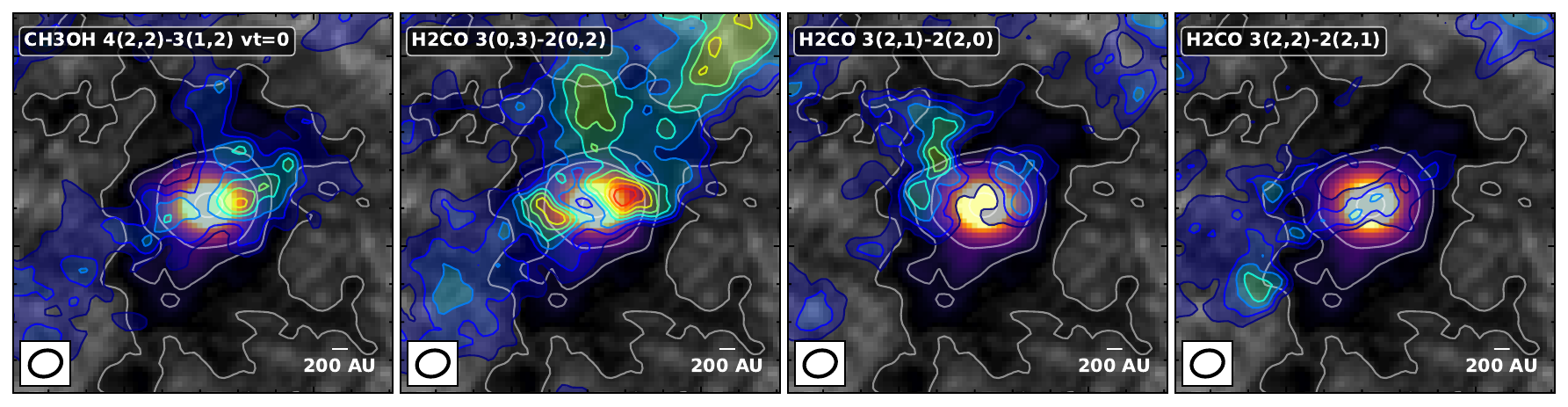}
\caption{Same as Fig.~\ref{fig:moment_maps} but shown for a zoomed-in field of view of $1'' \times 1''$ centred on the continuum peak of C2c1a(a).}
\label{fig:moment_maps_zoom}
\end{figure*}

Figure\,\ref{fig:moment_maps_zoom} shows a zoomed-in view of the integrated intensity maps presented in Fig.\,\ref{fig:moment_maps}, focusing on the central region around the continuum peak. This reduced field of view emphasises the spatial distribution of the molecular line emission relative to the compact core structure and allows a clearer assessment of the correspondence between line and continuum emission at the smallest resolved scales.

\end{appendix}

\end{document}